\documentclass{jpsj-suppl}
\usepackage{txfonts} 

\title{Tuning of the Charged Hadrons Multiplicities for Deep Inelastic Interactions in NEUT}

\author{Christophe \textsc{Bronner}$^{1}$ and Mark \textsc{Hartz}$^{1,2}$}

\inst{$^{1}$Kavli Institute for the Physics and Mathematics of the Universe (WPI), The University
of Tokyo Institutes for Advanced Study, University of Tokyo, Kashiwa, Chiba, Japan\\
$^{2}$TRIUMF, Vancouver, British Columbia, Canada}

\email{christophe.bronner@ipmu.jp}

\recdate{January 28, 2016}

\abst{We describe a procedure to tune the charged hadron multiplicities for deep inelastic events produced by the NEUT neutrino interaction generator. This tuning uses a model based on Koba-Nielsen-Olesen scaling, whose parameters are obtained by fitting multiplicity data from deuterium bubble chamber experiments. After tuning, the multiplicities of the events generated by NEUT are found to be in good agreement with the measurements from the bubble chamber experiments. }

\kword{Neutrino, deep-inelastic, NEUT, multiplicity tuning}

\begin{document}
\maketitle

\section{Introduction}
\subsection{Motivations}
Two of the main open questions studied by neutrino oscillation experiments are the neutrino mass hierarchy, and whether CP symmetry is conserved in those oscillations. In both cases, the experiments are trying to measure differences between neutrinos and anti-neutrinos, and so try to distinguish between interactions of neutrinos and anti-neutrinos in their detectors. In the case of experiments using water Cerenkov detectors, the separation between those interactions can only be statistical as the detectors are not magnetized. Neutrino enriched and anti-neutrino enriched samples can be built using the topological information of the events observed, including the number of reconstructed Cerenkov rings and Michel electrons. For such an approach to work, the distributions of those quantities have to be simulated properly by the simulation software used to do statistical analyses. For neutrinos of a few GeV, they are affected by the multiplicities of charged hadrons, in particular charged pions, produced in deep inelastic (DIS) interactions. In this paper we describe a method to tune the predictions of the neutrino interaction generator NEUT\cite{NEUT}, so that the events generated follow the charged hadrons multiplicity distributions measured by deuterium bubble chamber experiments\cite{BEBC, FNAL}.\\

\subsection{Deep inelastic events in NEUT}
In NEUT, events above the pion production threshold are generated differently depending on the value of the invariant mass of the hadronic system W. For values of W lower than 2 GeV/c$^{2}$, different resonant channels are considered (single pion, single kaon and single eta) as well as a DIS background called the `multi-pi' mode. This mode contains all the interactions for which W$<$2 GeV/c$^{2}$ and 3 hadrons or more are produced (all the hadrons outside of the ejected nucleon are assumed to be pions). Above 2 GeV/c$^{2}$, all the events are generated using PYTHIA 5.72\cite{Pythia}, this second region constitutes the `DIS' mode. The tuning described here is applied to the multi-pi and DIS modes.

\section{Tuning of the charged hadron multiplicities}
\subsection{Tuning procedure}
The tuning is done by applying a weight to the events generated by NEUT, which is a function of a certain multiplicity and of the value of W for each event. The bubble chamber experiments measured the charged hadron multiplicities, and this multiplicity is used to compute the weights for the DIS mode. Due to the way multiplicities are generated in NEUT for the multi-pi mode, the total hadron multiplicity is used instead for this mode. The weight is the ratio of the probabilities to obtain an event with a given multiplicity and W according to the bubble chamber experiments and to NEUT:

\begin{center}
  $\displaystyle{\text{Weight}=\frac{P_{Bubble \, chambers}(n, W)}{P_{NEUT}(n,W)}}$
\end{center} 

The probabilities according to the bubble chamber experiments are obtained by fitting the data of those experiments using a model, whereas the probabilities according to NEUT are obtained by generating a large number of events with this generator. The probabilities are evaluated separately for interactions of neutrinos and anti-neutrinos, and for interactions on protons and on neutrons. Finally, the weights are normalized so as to preserve the total cross-section as a function of W.

\subsection{Model used to fit bubble chamber experiments data}
The model used is based on KNO scaling\cite{KNO}, according to which the W dependance of the charged hadron multiplicity $P(n_{ch},W)$ can be fully taken into account through the average charged hadron multiplicity as a function of W, $<n_{ch}>(W)$:

\begin{equation}
  \displaystyle{P(n_{ch},W)=P(n_{ch},<n_{ch}>(W))}
\end{equation} 

The probability $P(n_{ch},W)$ can therefore be computed in two steps, as is done for example in the multi-pi mode of NEUT and in the AGKY model\cite{AGKY}. First, the average multiplicity for this W is computed using the empirical relation:
\begin{equation}
  \displaystyle{ <n_{ch}>(W)= A + B \times \ln(W^{2})}
\end{equation} 

Then the probability distribution of the multiplicity is obtained using a slightly modified version of KNO scaling (compared to standard KNO scaling, we have added a parameter $\alpha$ as introduced in \cite{Baranov}):
\begin{equation}
\displaystyle{P(n_{ch},W)=\frac{1}{<n_{ch}>-\alpha} \times f\left(\frac{n_{ch}-\alpha}{<n_{ch}>-\alpha}, C\right)},
\end{equation}
where we use for f the ``Levy function'' used in the AGKY model\cite{AGKY}. This gives a model with 4 parameters A,B,C and $\alpha$. The values of A and B are obtained by fitting the average multiplicity as a function of W in the bubble chamber data, whereas C and $\alpha$ are obtained by fitting the RMS versus the mean of the multiplicity distributions for the different W bins.

\subsection{Results of the fit}
\begin{figure}[tbh]
\begin{center}
\begin{minipage}{0.9\textwidth}
 \includegraphics[width=0.45\textwidth,clip]{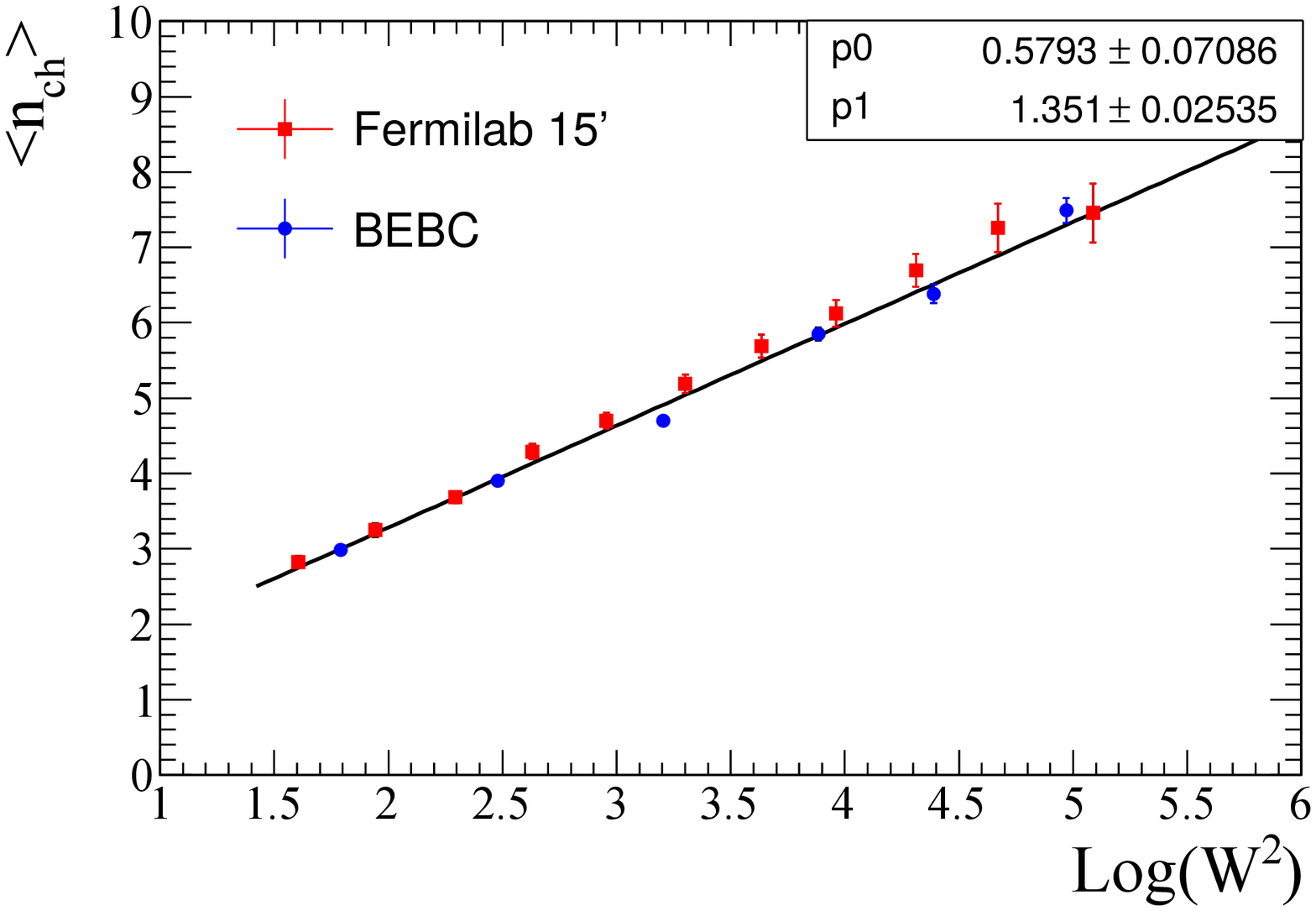} 
 \hfill
 \includegraphics[width=0.45\textwidth,clip]{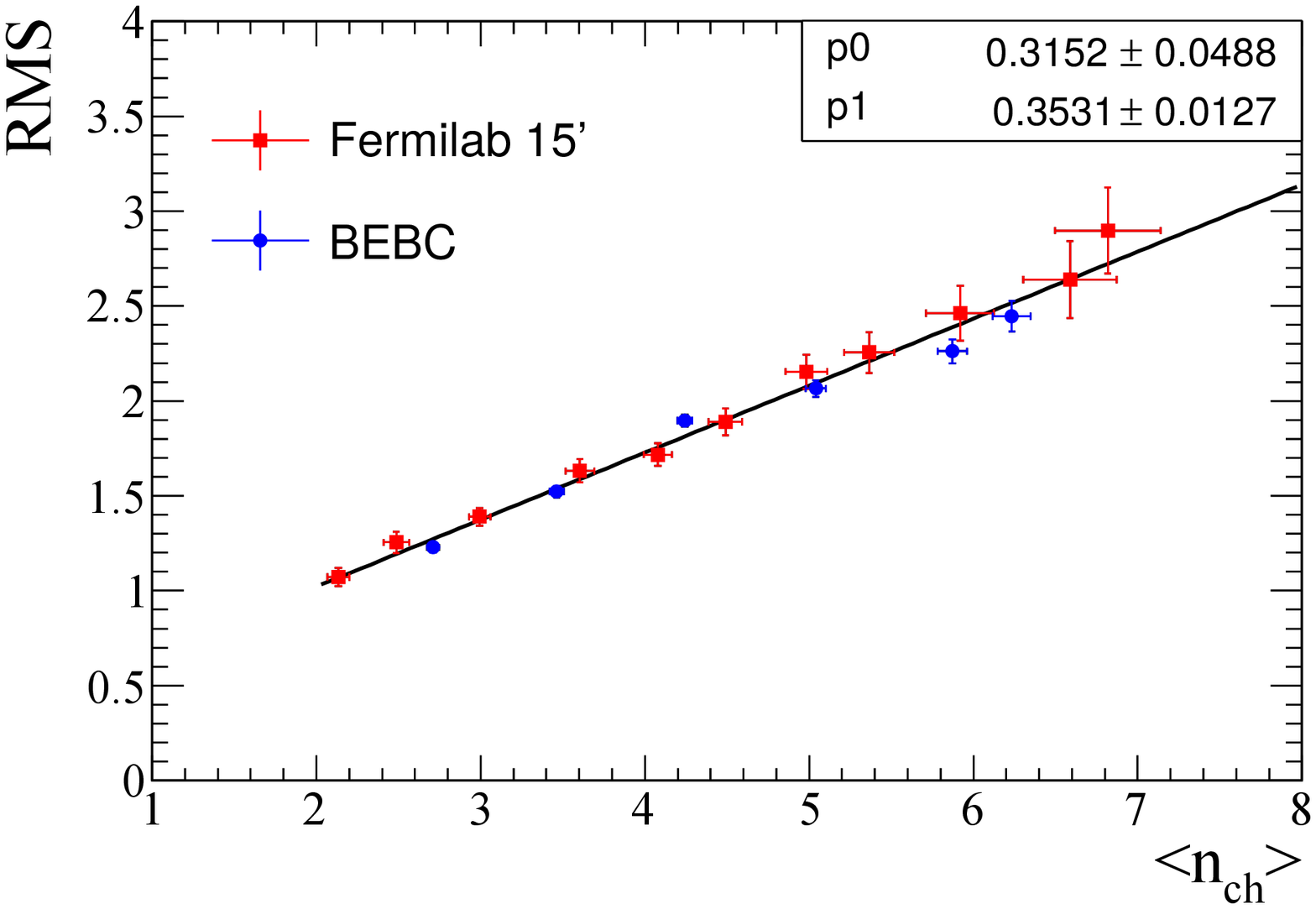}
\caption{Results of the fit of the bubble chamber data. Left plot: fit of the average charged hadron multiplicities as a function of W for interactions of neutrinos on protons. Right plot: fit of the RMS versus the mean of the multiplicity distributions for different W bins for neutrino interactions on neutrons.}
\end{minipage}
\label{FitResults}
\end{center}
\end{figure}

We use data from bubble chamber experiments on deuterium target, approximated as free neutrons and protons, to avoid the difficulties of modeling reinteractions in the nucleus. Based on the review of the different experiments done in \cite{PRC88}, we use data from two publications: one from the Fermilab 15' bubble chamber containing measurements for neutrino interactions\cite{FNAL}, and one from the Big European Bubble Chamber containing measurements for both neutrinos and anti-neutrinos\cite{BEBC}. Some of the fits are shown as examples in figure \ref{FitResults} (note that C and $\alpha$ are not the parameters directly obtained in the fit on the right plot, their values are deduced from the results of the linear fit), and the values of the parameters obtained in the different cases are summarized in table \ref{FitParams}. 

\begin{table}[tbh]
\caption{Values of the model parameters obtained by fitting data from bubble chamber experiments.}
\label{FitParams}
\begin{center}
\begin{tabular}{lcccc}
\hline
Parameter & $\nu$-proton & $\nu$-neutron  & $\overline{\nu}$-proton & $\overline{\nu}$-neutron\\
\hline
A & 0.58 $\pm$ 0.071& 0.35 $\pm$ 0.057 &0.41 $\pm$ 0.26& 0.80 $\pm$ 0.17 \\
B & 1.35 $\pm$ 0.025& 1.24 $\pm$ 0.02 &1.18 $\pm$ 0.088& 0.94 $\pm$ 0.065  \\
$\alpha$ & -0.58 $\pm$ 0.2& -0.89 $\pm$ 0.14& 0.68 $\pm$ 0.36& 0.38 $\pm$ 0.25 \\
C & 9.92 $\pm$ 0.90& 8.02 $\pm$ 0.58 &4.10 $\pm$ 0.75& 3.02 $\pm$ 0.47 \\
\hline
\end{tabular}
\end{center}
\end{table}

\section{Effect of the tuning}
\begin{figure}[tbh]
\begin{center}
\begin{minipage}{0.9\textwidth}
 \includegraphics[width=0.45\textwidth,clip]{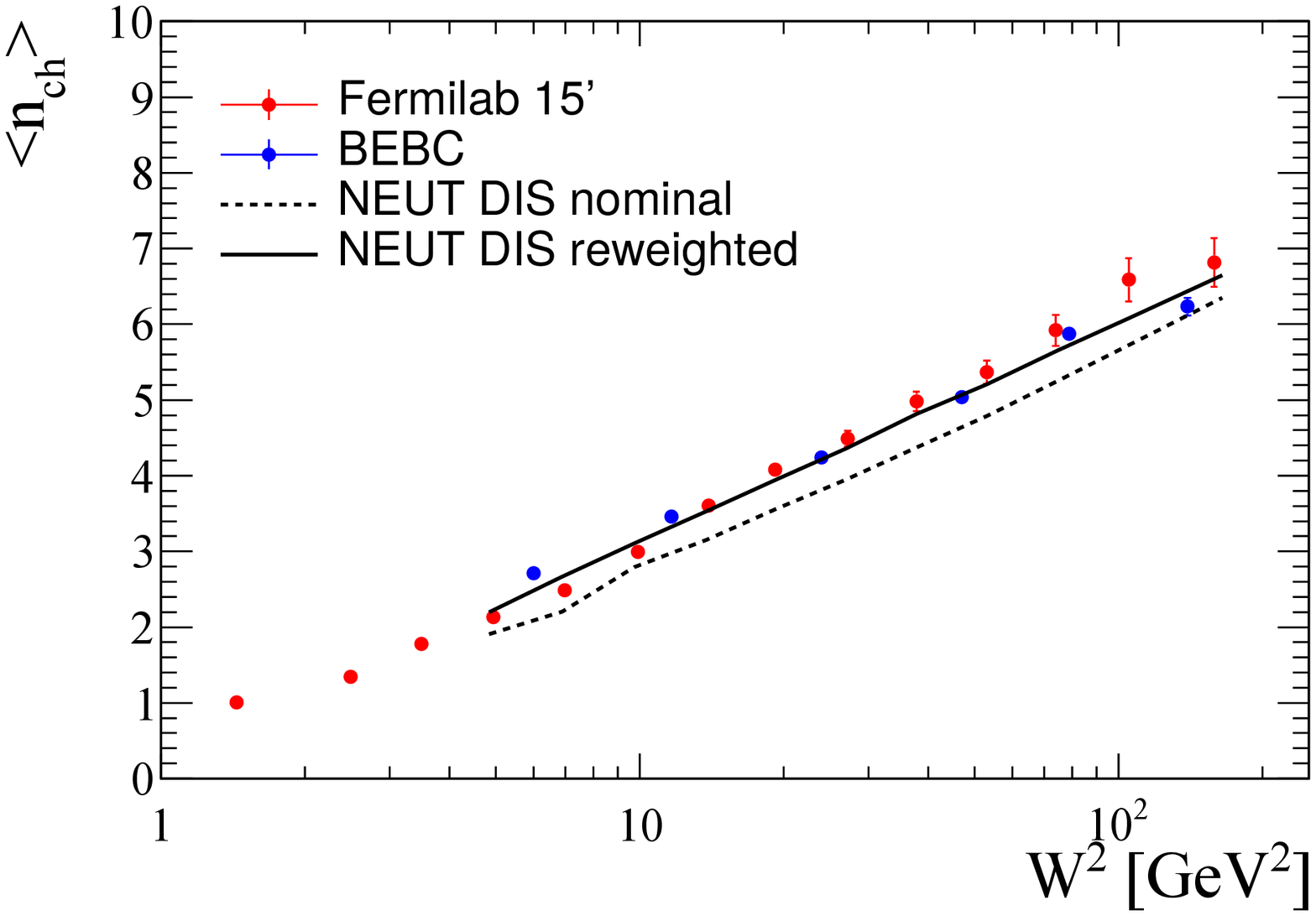} 
 \hfill
 \includegraphics[width=0.45\textwidth,clip]{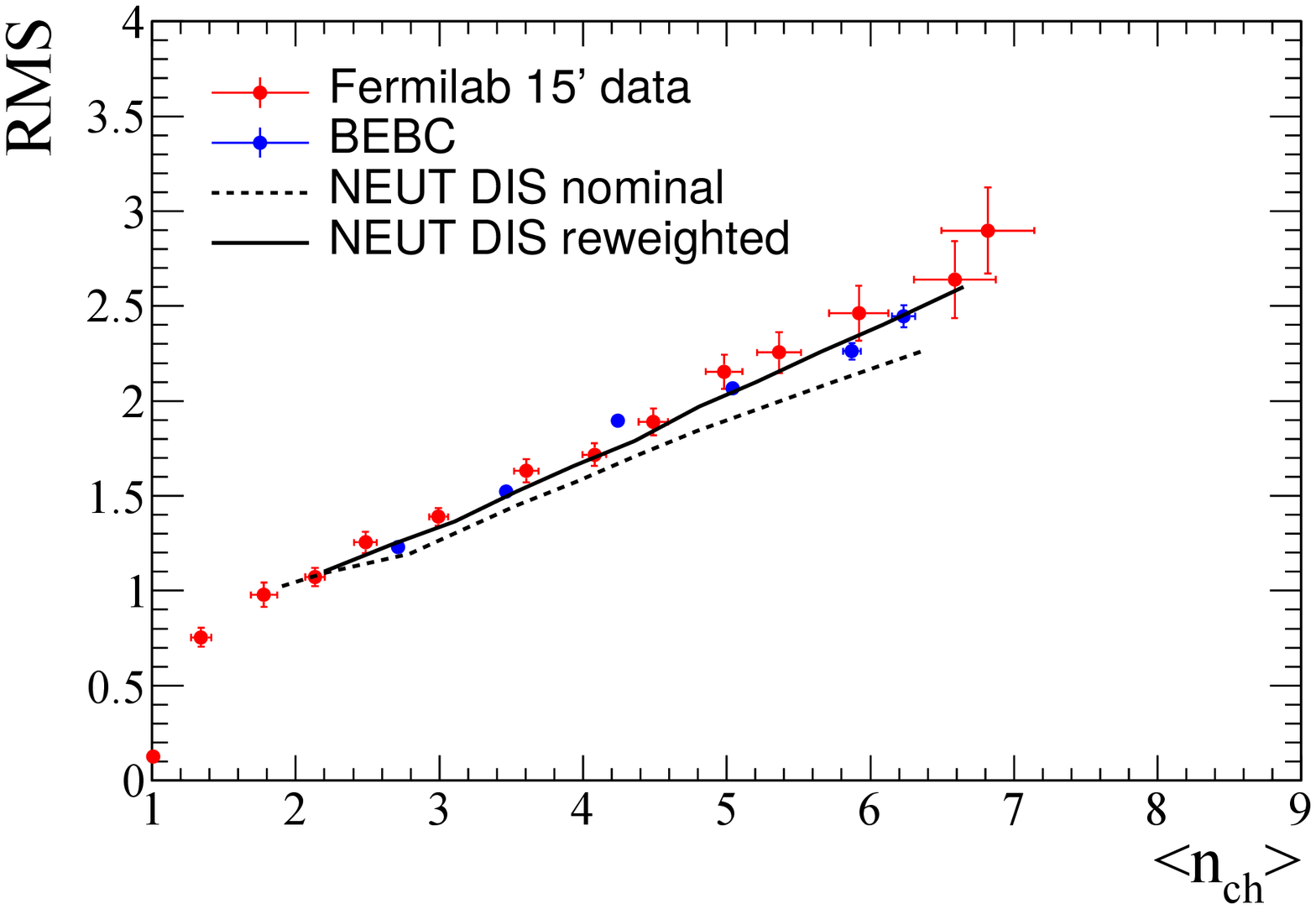}
\caption{Effect of the tuning for the DIS mode of NEUT, in the case of interactions of neutrinos on neutrons. The left plot shows the effect on the average charged hadron mulitplicity, and the right plot on the dispersion of those multiplicities.}
\end{minipage}
\label{TuningEff}
\end{center}
\end{figure}

A sample of NEUT events was generated according to the flux of atmospheric muon neutrinos and anti-neutrinos reaching the Super-Kamiokande detector\cite{SK}, to have a sample of events in a relevant range of hadronic invariant mass to test the tuning procedure. In the case of the DIS mode, the nominal predictions of NEUT were found to underestimate both the average and the dispersion of the charged hadrons multiplicities compared to the bubble chamber data. As can be seen wih the example of the interactions of neutrinos on neutrons shown on figure \ref{TuningEff}, the tuning allows to get good agreement with the data. 

\begin{figure}[tbh]
\begin{center}
\begin{minipage}{0.9\textwidth}
 \includegraphics[width=0.45\textwidth,clip]{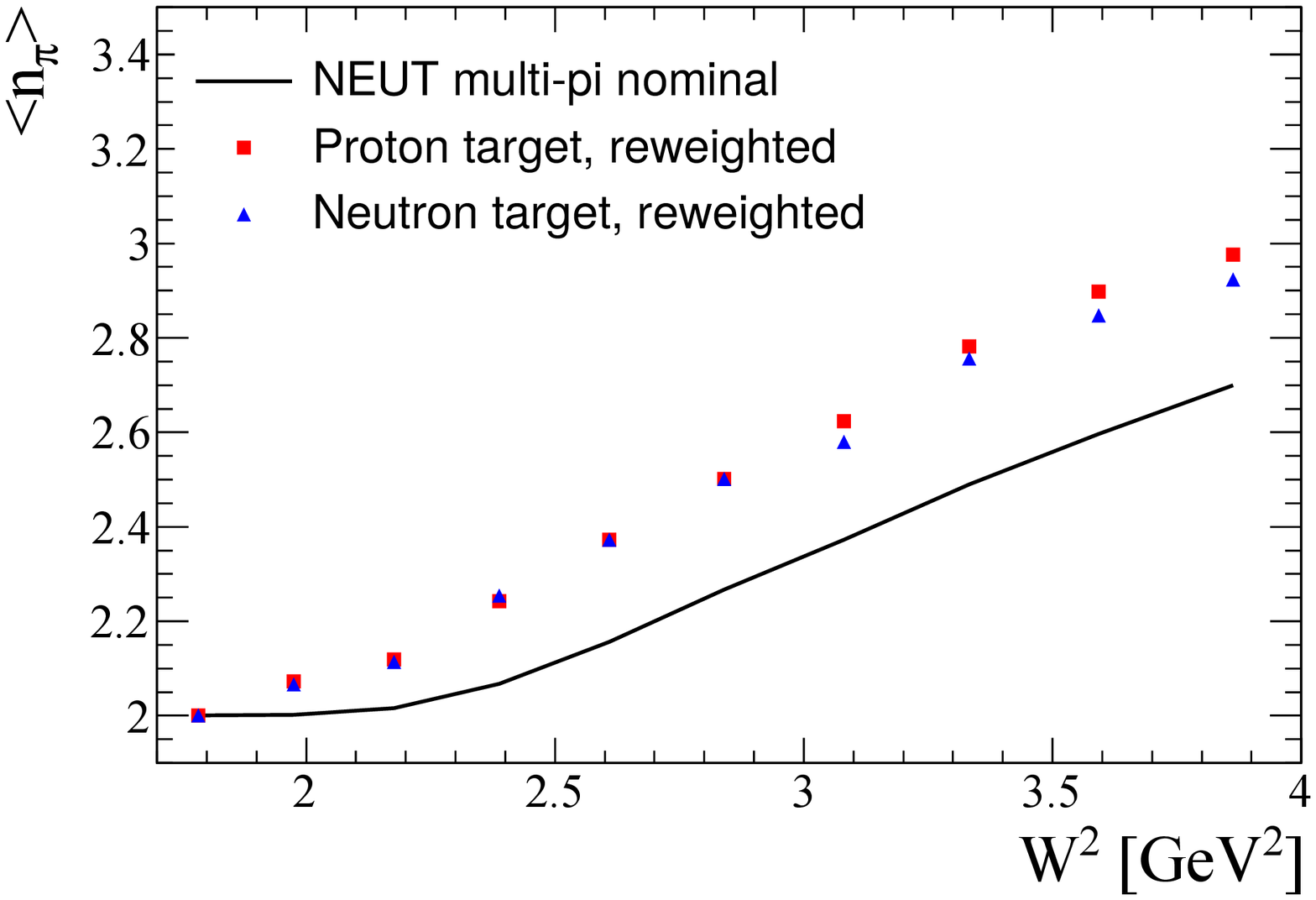} 
 \hfill
 \includegraphics[width=0.45\textwidth,clip]{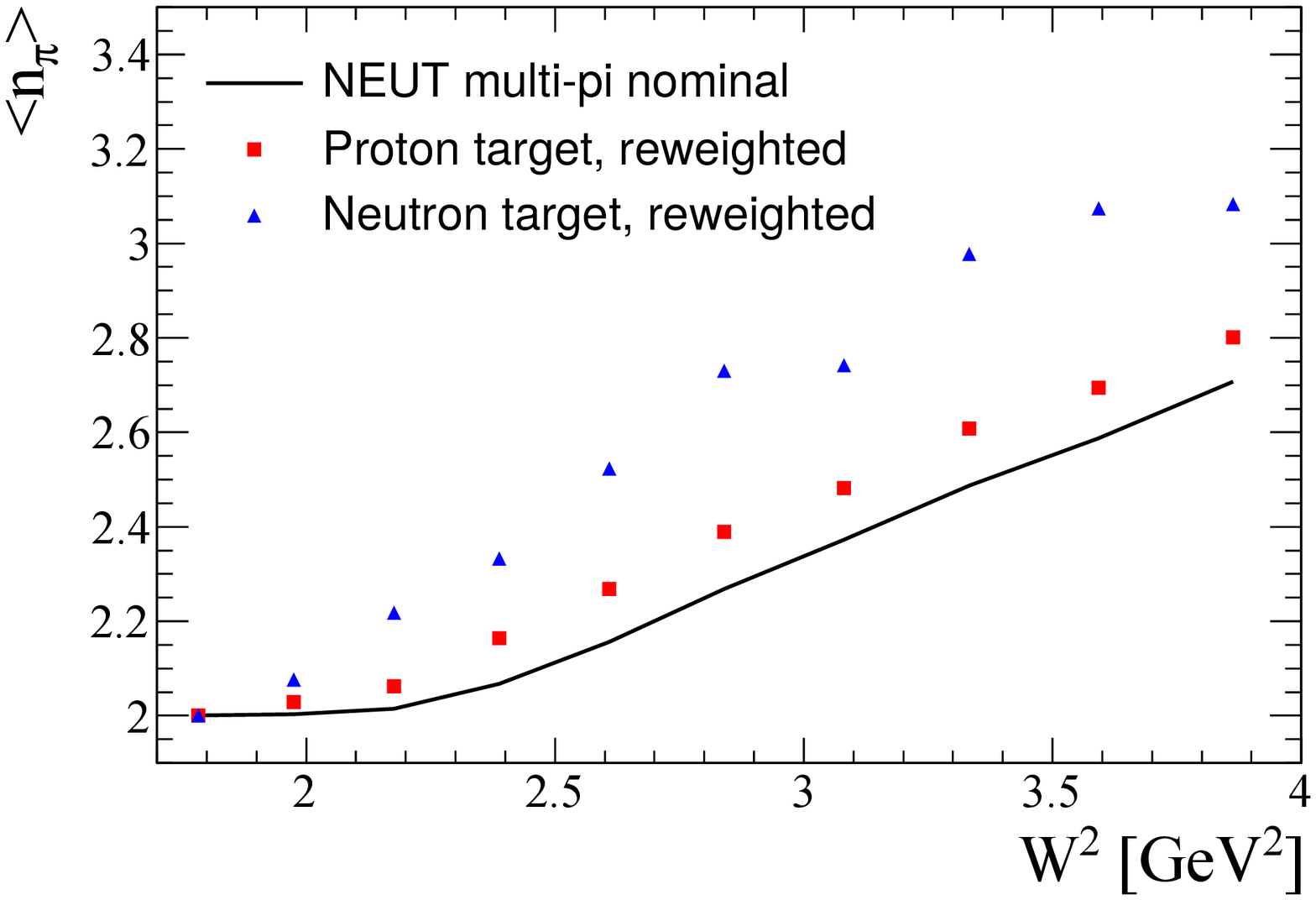}
\caption{Effect of the tuning on the average number of pions in events from the multi-pi mode of NEUT. Left plot corresponds to neutrino interactions, and right plot to anti-neutrino interactions.}
\end{minipage}
\label{TuningEffMP}
\end{center}
\end{figure}

In the case of the NEUT multi-pi mode, direct comparison to data is more difficult as it would require including the resonant events to which this tuning is not applied, and which have significant uncertainties of their own. We can however note that the tuning increases the average number of pions produced in events from this mode as can be seen on figure \ref{TuningEffMP} (from the definition of the multi-pi mode, all events have at least 2 pions). 

\section{Conclusion}
We described a tuning procedure of the charged hadron multiplicities for DIS events produced by the neutrino interaction generator NEUT. This procedure allows to reproduce the results of the multiplicity measurements performed by deuterium bubble chamber experiments. The final goal of this work is to use the model described here to propagate uncertainties on the hadron multiplicities to physics analyses, the next step will therefore be to evaluate the systematic uncertainties on this tuning. Those include the errors on the fitted values of the parameters, the differences between deuterium and free nucleons, as well as the hypothesis used in the case of the multi-pi mode to go from a model of the charged hadron multiplicities to a model of the multiplicities of all hadrons in DIS interactions.

\end{document}